# Electric field controlled magnetic exchange bias and magnetic state switching at room temperature in Ga doped α-Fe$_2$O$_3$ oxide

R.N. Bhowmik[*], and Abdul Gaffar Lone

Department of Physics, Pondicherry University, R.V. Nagar, Kalapet, Pondicherry-605014, India

[*]Corresponding author: Tel.: +91-9944064547; Fax: +91-413-2655734

E-mail: rnbhowmik.phy@pondiuni.edu.in

**Abstract**

We have developed a new magnetoelectric material based on Ga doped α-Fe$_2$O$_3$ in rhombohedral phase. The material is a canted ferromagnet at room temperature and showing magneto-electric properties. The experimental results of electric field controlled magnetic state provided a direct evidence of room temperature magnetoelectric coupling in Ga doped α-Fe$_2$O$_3$ system. Interestingly, (un-doped) α-Fe$_2$O$_3$ system does not exhibit any electric field controlled magnetic exchange bias shift, but Ga doped α-Fe$_2$O$_3$ system has shown an extremely high electric field induced magnetic exchange bias shift up to the value of 1120 Oe (positive). On the other hand, in a first time, we report the electric field controlled magnetic state switching both in α-Fe$_2$O$_3$ and in Ga doped α-Fe$_2$O$_3$ systems. The switching of magnetic state is highly sensitive to ON and OFF modes, as well as to the change of polarity of applied electric voltage during in-field magnetic relaxation experiments. The switching of magnetic state to upper level for positive electric field and to down level for negative electric field indicates that electric and magnetic orders are coupled in the Ga doped hematite system. Such material is of increasing demand in today for multifunctional applications in next generation magnetic sensor, switching, non-volatile memory and spintronic devices.

## 1. Introduction

The conventional spintronics devices uses spin-transfer torque technique, where spin-polarized current or magnetic field controls magnetic state and switching of magnetization [1]. This is high power consuming process. The main problem of the current or magnetic field controlled magnetization switching is the Joule heating effect during operation of the spintronics devices. Alternatively, electric field controlled switching of magnetization is an energy-efficient technique (less Joule heating effect) for the development of low power consumption spintronic devices with additional non-volatile functionality [2]. The electric field controlled magnetization state can be achieved in a special class of magneto-electrics, known as multiferroelectrics, where magnetization and electric polarization are strongly coupled in a crystal structure. Although ferroelectric order (needs empty $d$ shell ions) and magnetic order (needs partially filled $d$ shell ions) are mutually exclusive in nature in a crystal structure, some of the magnetic oxides ($BiFeO_3$, $TbMnO_3$, $CoCr_2O_4$) showed signature of multi-ferroelectric properties [3-4]. Among them, only few oxides have shown direct or indirect evidences of magneto-electric coupling where magnetic state has been controlled by electric field [5]. The concept of magneto-electric effect was emerged from the land mark discoveries of induced magnetism under electric field and induced electric polarization under magnetic field in a moving dielectric material [6]. Antiferromagnetic $Cr_2O_3$ was the first material that showed linear magneto-electric effect [7-8], which is not strong enough for room temperature application. The materials with room temperature magneto-electric properties are of increasing interest in the field of spintronics, non-volatile memory devices, data storage (electrically writing and magnetically reading), and sensor applications [9-11].

Recently, some hetero-structured materials, either naturally exist ($\beta$-$NaFeO_2$ [12]) or designed superlattices (($LuFeO_3$)$_9$/($LuFe_2O_4$)$_1$ [13], $Ti_{0.8}Co_{0.2}O_2$/$Ca_2Nb_3O_{10}$/$Ti_{0.8}Co_{0.2}O_2$ [14]) or theoretically predicted ($R_2NiMnO_6$/$La_2NiMnO_6$ [15]) exhibited electric field controlled magnetic exchange bias and switching at room temperature. These engineered magneto-electrics are mostly multi-layered structure of ferromagnetic(FM)/antiferromagnetic (AFM) films on ferroelectric (FE) substrate. $BiFeO_3$ is a well known single-phased oxide that established electric field controlled magnetic state near to room temperature. Recent reports [15-17] attributed the observed strong magnetoelectric coupling in the thin film of $BiFeO_3$ to large strain induced anisotropy generated at the interfaces of substrate and film. Otherwise, magnetoelectric coupling in bulk $BiFeO_3$ is very poor. Some ferrites, known as hexaferrites, with hexagonal structure (M-type: $BaFe_{10.2}Sc_{1.8}O_{19}$ [18] and $SrFe_{12}O_{19}$ [19], Z-type $Sr_3Co_2Fe_{24}O_{41}$ [20]) indicated electric field controlled magnetic state at room temperature and in most of the cases at lower temperatures [21]. At this point, we mention that derivatives of hematite ($\alpha$-$Fe_2O_3$), which is an anisotropic electrical insulator, could be interesting for developing material with electric field controlled magnetic switching properties. Some of the hematite derived systems ($GaFeO_3$, $FeTiO_3$) in orthorhombic phase showed magnetoelectric properties at lower temperatures [22-24] and some compositions in thin film form (e.g., Mn doped $Ga_{0.6}Fe_{1.4}O_3$:Mg [25] and $GaFeO_3$ [26]) showed magnetoelectric properties at room temperature. Further, hematite ($\alpha$-$Fe_2O_3$) derivatives in corundum (rhombohedral) structure offers electrically polar and room temperature multiferroic materials [27-28].

In an attempt of searching new magnetoelectrics, we have taken an extensive research program to develop materials with room temperature ferromagnetism and electric field controlled magnetic state based on rhombohedral phase of metal doped hematite system. We developed Ga doped hematite system in rhombohedral structure ($\alpha$-$Fe_{2-x}Ga_xO_3$) with canted ferromagnetic state at room temperature and good signature of ferroelectric polarization [29-31], although electric field dependence polarization curves are not completely free from leakage due to relatively high conductivity of the samples [32-33]. In this work, we establish Ga doped $\alpha$-$Fe_2O_3$ system as a new magnetoelectric material, where magnetic state at room temperature is controlled by external electric field and switching of magnetic state is sensitive to ON and OFF modes, as well as to the change of polarity of applied electric field.

## 2. Experimental

We have prepared the Ga doped $\alpha$-Fe$_2$O$_3$ ($\alpha$-Fe$_{2-x}$Ga$_x$O$_3$: x = 0.2-1.0) system by mechanical alloying of the fine powders of $\alpha$-Fe$_2$O$_3$ and $\beta$-Ga$_2$O$_3$ oxides. The alloying time was increased up to 100 h depending on the composition. We determined structural phase evolution of the alloyed samples using X-ray diffraction (XRD) pattern, recorded using Cu K$_\alpha$ radiation ($\lambda$ = 1.54054 Å). The samples with higher Fe content (x = 0.2) produced single phased structure by mechanical alloying itself. The samples with higher Ga content (x = 0.8) do not produce single phased structure by mechanical alloying alone and it needed special heat treatment under vacuum, as described in [29, 31]. To maintain identical heat treatment condition, we directly heated the mechanical alloyed powder at 800 $^0$C under high vacuum ($10^{-5}$ mbar) for 2-6 h with fast heating and cooling rate during change of temperature. The single phased rhombohedral structure (space group R$\overline{3}$C) was confirmed from XRD pattern and supported by Micro-Raman spectra (Fig. 1) for the samples used in this work with compositions $\alpha$-Fe$_{1.8}$Ga$_{0.2}$O$_3$, $\alpha$-Fe$_{1.6}$Ga$_{0.4}$O$_3$ and $\alpha$-Fe$_{1.2}$Ga$_{0.8}$O$_3$, respectively. The band at around 1320 cm$^{-1}$ (magnon-phonon mode) in Micro-Raman spectra confirmed a strong spin-lattice coupling in rhombohedral structure of Ga doped hematite system. The structural and magnetic properties for some of the Ga doped hematite samples in rhombohedral phase have been discussed in earlier works [29, 31-32]. On the other hand, mechanical alloying and subsequent vacuum annealing were not enough to produce single phased rhombohedral structure for the composition $\alpha$-FeGaO$_3$. The mechanical alloyed powder of this composition was heated at 1250 $^o$C in air to stabilize orthorhombic structure. The orthorhombic phase was transformed into rhombohedral phase by mechanical milling of the heated sample up to 50 h and the rhombohedral phase was refined by vacuum annealing of the sample at 700 $^0$C (2 h).

The vacuum annealing of milled sample at 800 $^0$C indicated re-appearance of orthorhombic phase. In this work, we used the rhombohedral phased sample (Ga10MM50V7) that was subjected to 50 h milling and vacuum annealed at 700 $^0$C (2 h). A brief description of the samples (Ga02MA100V8, Ga04MA50V8, Ga04MA100V8, Ga08MA25V8, Ga10MM50V7) and their structural information are provided in Table 1. The prepared material can be considered as a solid solution of Ga atoms into α-$Fe_2O_3$ structure, where non-magnetic Ga atoms have been dissolved in the lattice sites of magnetic Fe atoms. Variation of the lattice parameters in the samples is attributed to the differences in grain size and incorporation of $Ga^{3+}$ with smaller ionic radius (0.62 Å) into the lattice sites of $Fe^{3+}$ ions with larger radius (0.645Å). The notable feature is that peak intensity of IR-active $E_u$(LO) mode of lattice vibration (at about ~ 667 cm$^{-1}$) significantly increased and displaced to higher wave number for Ga10MM50V7 sample. The increase of the intensity of IR-active $E_u$(LO) mode reflects an increasing local disorder in the lattice structure of hematite at higher Ga content. The displacement of the peaks to higher wave number is related to the decrease of cell parameter in the lattice structure of hematite at higher Ga content. A pellet shaped sample of typical dimension 3 mm x 2 mm x 1 mm was placed between two thin Pt sheets, which were connected to 2410-C meter using thin Pt wires for applying dc electric voltage during measurement of dc magnetization with magnetic field and time by using vibrating sample magnetometer (LakeShore 7404, USA). The sample sandwiched between Pt electrodes was placed on the flat surface of the sample holder (Kel-F) and tightly fixed by Teflon tape. The proper electrical contact has been checked from identical values of current on reversing the applied voltage at ±5 V.

## 3. Experimental results

Fig. 2(a) shows the M(H) loop for hematite sample at room temperature, measured under different values of applied electric voltage (0-300 V). The inset of Fig. 2(a) (magnified loop at 0 V and 300 V) confirms the absence of any electric field induced shift of M(H) loop in hematite sample. The time dependence of magnetization in the presence of 5 kOe magnetic field (in-field magnetic relaxation) in hematite sample (Fig. 2(b)) shows a typical character of an antiferromagnet or canted antiferromagnetic system [34-35]. Interestingly, magnetic state of hematite sample, irrespective of the increase or decrease of in-field magnetization with time (Fig. 2(c-d)), is switchable by external electric voltage. The magnetic state is highly sensitive to the change of polarity, as well as to ON and OFF switching modes of the external electric voltage. One can see an instant magnetization jump to higher state (lower state) by switching ON the electric voltage +100 V (-100 V) and returned back to original magnetic state by switching OFF the electric voltage. The magnetic switching is repeatable and the change of magnetization is nearly 0.87-1.00 % by applying electric voltage at $\pm100$ V.

On the other hand, electric field controlled magnetic properties in Ga doped hematite system are remarkably different from hematite sample. We present details experimental results for the compositions $\alpha$-Fe$_{1.8}$Ga$_{0.2}$O$_3$ (Ga02MA100V8), $\alpha$-Fe$_{1.6}$Ga$_{0.4}$O$_3$ (Ga04MA50V8, Ga04MA100V8), $\alpha$-Fe$_{1.2}$Ga$_{0.8}$O$_3$ (Ga08MA25V8) and $\alpha$-FeGaO$_3$ (Ga10MM50V7), respectively. The M(H) loops of Ga02MA100V8 sample (Fig. 3(a)) showed a shift by increasing the applied voltage up to 200 V. Fig. 3(b) shows that M(H) loop of the sample shifted along the direction of positive magnetic field and negative magnetization for applying both positive and negative bias voltage. The negative bias voltage has little effect on changing the nature of the loop shift. But, M(H) loop at -200 V has slightly shifted towards negative magnetic field direction in comparison to the loop measured at +200 V. The magnetic exchange bias field (H$_{exb}$) has been calculated from the shift of the center of the M(H) loop under electric voltage with respect to the center of the M(H) loop at 0 V. Fig. 3(c) shows that magnetic exchange bias field (H$_{exb}$) of the Ga02MA100 sample can be increased up to + 1226 Oe for applied voltage +200 V and H$_{exb}$ was found up to +1190 Oe for applied voltage -200 V. On the other hand, magnetic state of the Ga02MA100V8 sample during in-field magnetic relaxation process is instantly jumps to higher magnetization state by applying +100 V (Fig. 3(d) and to lower magnetization state by applying -100 V (Fig. 3(f), in addition to the natural magnetic relaxation process (decreasing trend) of the spins with time. Despite the fact of a rapid change of magnetization with time during ON and OFF modes of electric voltage, the in-field magnetic relaxation confirms a step-wise increment/decrement of magnetization of the sample with the step-wise (100 V) increment/decrement of applied voltage up to ±500 V. This resulted in the change of magnetization by 0.96% and 0.60 % for the applied voltage +500 V and -500 V, respectively with respect to 0V reference level.

Fig. 4(a) shows the M(H) loop for Ga04MA100V8 sample, measured in the magnetic field range ±16 kOe and in the presence of constant electric voltage 0 V to + 200 V. The M(H) loops under positive electric voltage shifted towards the positive magnetic field axis and negative magnetization direction in comparison to the M(H) loop measured at 0 V. The electric field induced shift of M(H) loop is clearly visible from the plot within small magnetic field range (Fig. 4(b)). It may be noted that magnetization vectors in the M(H) curve reversed upon reversing the magnetic field directions. Fig. 4(c) shows that shift of the M(H) loop certainly depends on the polarity of applied electric voltage. The M(H) loop shifted towards positive magnetic field direction upon application of + 200 V (positive). The loop shift under negative voltage (-200 V) is not exactly the mirror image of the loop at + 200 V with reference to the loop at 0 V. On reversing the electric field from + 200 V to -200 V, the magnetization shifted towards negative magnetic field and positive magnetization directions when measured at -200 V in comparison to the M(H) loop measured at + 200 V. However, both the M(H) loops under opposite polarity of electric voltage remained in the same positive side of the magnetic field with respect to the loop at 0 V. This results in a decrease of positive magnetic exchange bias field by nearly -75 Oe during measurement at - 200 V (with $H_{exb} =$

+267 Oe) with reference to the exchange bias field ($H_{exb}$ = +392 Oe) during measurement at + 200 V. The results are similar to that observed in the synthetic multiferroelectric film, consisting of $La_{0.67}Sr_{0.33}MnO_3$ (ferromagnet) and BaTiO3 (ferroelectric) [35]. Fig. 4(d) shows a rapid increase of the magnetic exchange bias field at the initial increment (up to 20 V) of electric voltage and subsequently slowed down at higher electric voltages to achieve the value that falls in the range of magnetic coercivity (362 ± 10 Oe) for Ga04MA100V8 sample. The magnetic coercivity ($H_C$) has been calculated from the average of the values in positive and negative magnetic field axis of the M(H) loop. The Ga04MA50V8 sample also exhibited similar kind of electric field controlled M(H) loops (Fig. 5(a)) and a rapid increase of magnetic exchange field ($H_{exb}$) for applied electric field < 20 V and then slowed down to achieve a saturation value for electric voltage above 50 V (inset of Fig. 5(a)). The magnetic coercivity ($H_C$) of the sample at higher electric voltages approaches to the value 353 Oe observed at 0 V, whereas Ga04MA50 sample has achieved the $H_{exb}$ up to + 370 Oe for electric voltage at + 200 V. Fig.5(b) shows that M(H) loop shift of the Ga04MA50 sample depends on polarity of the applied electric voltage. The M(H) loop shifts towards the negative magnetic field axis when measured at -200 V in comparison to the M(H) loop measured at electric voltage at + 200 V. Subsequently, magnetic exchange bias shift for negative electric voltages (-10 Oe at -100 V and - 40 Oe at -200 V) is significant with respect to the values at positive voltages (+ 100 V and + 200 V). The $H_{exb}$(V) curve for positive and negative electric field variation is nearly symmetric about a reference line (inset of Fig. 5(b)), which lies in between $H_{exb}$(+V) and $H_{exb}$(-V) curves, and of course, not with respect to $H_C$(0 V) line.

We have examined the electric field controlled magnetic state at room temperature for two samples of $\alpha$-$Fe_{1.6}Ga_{0.4}O_3$ (Fig.6 for Ga04MA100V8 and Fig.7 for Ga04MA50V8) through in-field magnetic relaxation experiment (time dependence of magnetization at constant magnetic field 5 kOe). The in-field magnetic relaxation behaviour in both the samples is more or less similar in character (i.e., M(t) decreases in the presence of 5 kOe field) as in hematite sample (Fig. 2(b)). Fig. 6(a) shows the example of magnetic relaxation of Ga04MA100V8 sample in the presence of 5 kOe and at 0 V. The magnetization in both the samples is highly sensitive and switchable under ON-OFF modes and reversal of the polarity of applied electric voltage, as measured at + 100 V (Fig. 6(b), Fig. 7(a)), at -100 V (Fig. 6(c), Fig. 7(b)), at cyclic order of the polarity change with sequence 0 V→+ 100 V→ 0 V → -100 V (Fig. 6(d), Fig. 7(c)). As shown in Fig. 7(d-e), magnetization in both the samples showed a sudden jump in response to the change of applied voltage either from ON to OFF or OFF to ON modes. A complete reversal of the magnetization vectors (negative to positive or vice versa) is not observed macroscopic level upon reversing the polarity of electric voltage during in-field magnetic relaxation of spins (coloured symbol), but the samples instantly jumped to higher magnetic state at the time of applying positive voltage ($\Delta$M ~ 1.36 %) and to lower magnetic state at the time of negative voltage ($\Delta$M ~ 1.14 %) with respect to magnetization state at 0 V. The change of magnetization was calculated using the formula $\Delta$M (%) = $\frac{(M(V)-M(0))*100}{M(0)}$. The in-field magnetization being in the meta-stable state relaxed slowly with time even in the presence of electric field, irrespective of positive or negative sign, towards achieving the magnetization state at 0 V. However, there exists a gap ($\Delta$M ~ 0.13-0.26 %) between the relaxed magnetic state just before and after switching the ON/OFF modes of electric voltage. It indicates that switched magnetic state after relaxation is different from the relaxed state at 0 V. The existence of magnetization gap during M(t) measurement is irrespective of the magnitude, polarity and cycling of electric voltage. The relaxation of magnetization under simultaneous presence of electric and magnetic fields may be related to the kinetic energy transfer of charge-spin carriers. Fig. 6(e-f) showed that the relaxation rate is relatively fast at higher electric voltage ($\pm$ 200 V), where in-field magnetization under 5 kOe relaxed rapidly to the magnetization state either at 0 V preceding to the application of electric voltage or even continued to relax. Such rapid relaxation of magnetization under high electric field may be affected by magneto-conductivity that cannot retain the switched magnetization state for long time. Fig. 7(f) shows the in-field magnetic relaxation where applied electric voltage has been increased step-wise (size 50 V) up to + 500 V for every 50 s interval during measurement time up to 600 s. This experiment confirms a systematic increase of magnetization (magnetic spin order) with the increase of positive electric voltage. Fig. 7(g) shows that $\Delta M$ of the base and peak values of M(t) data under electric voltage continuously increased with the increase of electric voltage in comparison to the M(t) data at 0 V (during first 50 s of the measurement). The $\Delta M$ is found to be 2.51 % and 1.23 % for the peak and base values in the presence of electric voltage 500 V during last 50 s of the measurement time. The gap between $\Delta M$ (%) at peak and base line increased with applied voltage, which could be related to relaxation of the induced magnetization at higher voltage.

The composition $\alpha$-$Fe_{1.2}Ga_{0.8}O_3$ (Ga08MA25V8) also exhibited electric field controlled magnetic exchange bias shift (Fig. 8) and magnetic state switching (Fig. 9). Similar to other compositions of Ga doped hematite system, M(H) loop of the Ga08MA25V8 sample (Fig. 8(a)) shifted towards the positive magnetic field axis and negative magnetization direction when measured under positive electric voltage in comparison to the M(H) loop measured at 0 V. The M(H) loop shift also depends on the polarity of applied electric voltage (Fig. 8(b)). The M(H) loop under negative voltage (-200 V) has shifted towards positive magnetic field direction in comparison to the loop measured under + 200 V (positive). This results in a decrease of magnetic exchange bias from +85 Oe during measurement at + 200 V to +76 Oe during measurement at - 200 V. The inset of Fig. 8(b) shows that magnetic coercivity of the Ga08MA25V8 sample is limited within the range 210 ±5 Oe for applying the electric voltage in the range 0 V to 300 V. However, magnetic exchange bias shift increases with the increase of applied electric field and reached to + 152 Oe for the M(H) loop that was measured under electric voltage +300 V. Fig. 9 shows the in-field magnetic relaxation at 5 kOe magnetic field of the Ga08MA25 sample under different modes of the application of electric voltages. The magnetic state of the sample is highly sensitive and switchable under the ON-OFF modes and reversal of the polarity of applied electric voltage with respect to the in-field magnetic relaxation state at 0 V (Fig. 9(a)). The switching of magnetic state of the sample has been tested by a cyclic change of electric voltage with a sequence $0\ V \rightarrow + 200\ V \rightarrow 0\ V \rightarrow + 200$ V (Fig. 9(b), $0\ V \rightarrow - 200\ V \rightarrow 0\ V \rightarrow - 200$ V (Fig. 9(c)), $0\ V \rightarrow + 200\ V \rightarrow 0\ V \rightarrow - 200 \rightarrow$ $0\ V$ (Fig. 9(d)), and a step-wise increment of voltage up to + 500 V and -500 V separately with step size 100 V (Fig. 9(f))), at cyclic order of the polarity change with sequence $0\ V \rightarrow +$ $100\ V \rightarrow 0\ V \rightarrow -100\ V$ (Fig.(e)). The concrete information from Fig. 9 is that magnetic state of the sample is highly switchable under the application of electric field. The magnetic state instantly jumps to higher magnetization level in response to positive voltage and to lower magnetization level in response to negative voltage, irrespective of the intermediate in-field magnetic relaxation of the sample. The magnetization in the presence of electric voltage with respect to 0 V line systematically increases/decreases by the increment/decrement of the voltage up to +500 V/-500 V with step size 100 V and the change was up to 2 % (Fig. 9(f)). Fig. 10(a) shows that rhombohedral structured sample (Ga10MM50V8) of the composition $\alpha$-$Fe_{1.8}Ga_{0.2}O_3$ is a weak ferromagnet at room temperature. This sample in the rhombohedral phase also exhibits (Fig. 10(b)) electric field induced magnetic exchange bias shift (up to at 14 Oe at +200 V), although the ferromagnetic order in this composition has been diluted due to more amount of non-magnetic Ga content and lattice disorder. We have shown in Fig. 10(c) that the exchange bias shift for different composition of $\alpha$-$Fe_{2-x}Ga_xO_3$ system is close or less than the values of magnetic coercivity of the corresponding samples. Both these magnetic parameters have decreased with increase of Ga content in hematite structure.

## 4. Discussion

The electric field controlled magnetic state has been reported mainly in thin-films of FM semiconductors, multiferroics, and multi-layered structure consisting of ferroelectric (FE) and ferromagnetic (FM) materials [1-2]. It has been proposed that change of charge carrier density by electric field changes the magnetic exchange interactions in FM semiconductor. On the other hand, charge (electric polarization) and spin (magnetization) coupling by applied electric field can modify the magnetic state in multiferroic material. The electric field controlled magnetic exchange bias was first explored at the heterostructured interface of $Cr_2O_3$ $(Co/Pt)_3$ [37], where electric field controlled magnetism was attributed to strain-mediated magnetoelectric coupling at the heterojunction of FM and FE layers [36, 38]. The present Ga doped hematite system is not in thin film form and the observed electric field induced magnetic exchange bias shift can be excluded from strain mediated electro-magnetic coupling effect. Also, magnetic coercivity in the present samples with specific Ga content is nearly independent of applied electric field, unlike the increase of magnetic exchange bias shift with applied electric voltage. This means the magnetic anisotropy, related to spin–orbit interaction of electrons or interfacial anisotropy as proposed for magnetic tunnel junctions in FM semiconductors [1], is practically insensitive to external electric field in Ga doped hematite system. Hence, the electric field induced magnetic state in rhombohedral structure of Ga doped hematite system over a wide range of Ga content could be originated from different sources. Since the material is new and there is not any straight-forward mechanism or theoretical model available for explaining the electric field induced magnetic state, we understand the observed electric field induced magnetic state using models proposed for different systems and seems to be reasonable for the Ga doped hematite system. In the absence of a significant role of the strain mediated interfacial coupling and spin-orbit interactions for showing electric field controlled magnetic state in Ga doped hematite samples, we focus on electric field dependent perturbation in the magnetic spin order at the domain walls as proposed for many artificially designed multi-layered materials [38-39].

We noted that layer kind lattice structure of material, whether single phased ($BiFeO_3$) [16] or multi-phased ($BaTiO_3$/$La_{0.67}Sr_{0.33}MnO_3$ [36]), seems to be more sensitive for showing electric field controlled magnetic exchange bias. R. Moubah et al. [40] directly captured the image of coupled FE and AFM domains in multiferroic BiFeO3 single crystal. It was observed that several AFM domains coexist inside one single FE domain. Skumryev et al. [41] explained the magnetization reversal and magnetic exchange bias (EB) shift in FM Ni81Fe19 (Py) film deposited on AFM multiferroic (LuMnO3) single crystal as the effect of electric-field driven decoupling between FE and AFM domain walls. They proposed the existence of clamped and unclamped AFM domain walls (AF-DWs) at the interfaces with FE domains. A coupling between clamped AFM-DWs and FE domains exerts electric field controlled torque on FM moments, where as unclamped AF-DWs do not play significant role in the electric field controlled magnetic switching. In fact, electric field induced magnetic exchange bias field in our material (reached up to 1220 Oe at 200 V for Ga02MA100V8 sample) is remarkably large in comparison to the reported values (~ 100-300 Oe) in hetero-structured multiferroic films [37, 41]. However, a similar variation of exchange bias shift on increasing the applied electric voltage was noted in the Ga doped hematite samples. The switching of (canted)ferromagnetic state by electric field indicates the creation of additional magnetic exchange interactions in the system. It may be mentioned that hematite ($\alpha$-$Fe_2O_3$) does not show any signature of ferroelectric properties and electric field induced magnetic exchange bias. On the other hand, Ga doped hematite system exhibited electric field induced magnetic exchange bias effect. Although leakage of polarization is not completely overcome due to high conductivity of the samples, but existence of ferroelectric polarization in Ga doped hematite samples with rhombohedral structure has been realized. However, there is a further scope of improvement of ferroelectric polarization in present material by preparing in thin film form, as described in some thin films of Ga doped hematite in orthorhombic phase [25-26]. The important point to be noted is that electrical polarization in the polycrystalline α-$Fe_{1.6}Ga_{0.4}O_3$ samples [33] is responded to magnetic field and polarization curve is well comparable quantitatively and qualitatively to that in Mn doped orthorhombic structured $Ga_{0.6}Fe_{1.4}O_3$:Mg thin films [25]. Even, the shape of the polarization curve in our Ga doped hematite sample is far better than that observed in M-type hexaferrite $SrFe_{12}O_{19}$ [19], an emerging multi-ferroelectric material. At the same time, magnetic coercivity decreased with the increase of Ga content in hematite structure. These features indicated a modified magnetic structure in Ga doped hematite system [29, 31-32].

We propose a schematic diagram (Fig. 11) to understand the electric field controlled magnetic spin order (exchange bias and switching) in Ga doped hematite system. As sketched in Fig. 11(a), the hematite system in rhombohedral structure with $R\bar{3}C$ space group is magnetically multi-layered spins structure, where in-plane $Fe^{3+}$ spins form FM order and $Fe^{3+}$ spins in alternating planes (say, A and B) along off-plane direction form AFM order by superexchange interactions ($Fe^{3+}_A$-$O^{2-}$-$Fe^{3+}_B$). The weak ferromagnetism arises due to canting among AFM aligned spins by Dzyaloshinskii–Moriya (DM) interactions [$\sim \vec{D}.(\vec{S_n} \times \overrightarrow{S_{n+1}})$]. The replacement of magnetic Fe atom by non-magnetic Ga atom increases magnetic non-equivalence between A and B planes. The enhancement of spin-lattice coupling and small atomic displacement within rhombohedral structure of Ga doped hematite are indicated from Micro-Raman spectra. The dielectric peak in the temperature regime (290 K-310 K), where $Fe^{3+}$ spins started flipping from in-plane direction (canted FM state) to out of plane direction [32, 42], indicated (canted) spin order induced magneto-electric coupling in Ga doped hematite system. We restrict our discussion for the origin of electric polarization in Ga doped hematite system based on the models of spin induced ferroelectricity [5, 24]. The spin induced ferroelectricity/multiferroic properties (electric field controlled magnetism) has been observed at lower temperature in a family of orthoferrites $RFeO_3$ (R = Gd, Tb, Dy) [43]. In such spin canted systems, $\vec{S_i} \times \vec{S_j} \neq 0$ where $S_i$ and $S_j$ represent spins at two canted spin sites. According to inverse Dzyaloshinskii-Moriya (DM) model [18, 39], the antisymmetric spin exchange interactions $\hat{e}_{ij} \times (\vec{S_i} \times \vec{S_j})$, where $\hat{e}_{ij}$ is the unit vector connecting spins at sites i and j, can produce macroscopically observable polarization ($\vec{P_{Ij}}$) under the influence of relativistic spin–orbit coupling, despite the lack of non-centrosymmetric displacement of cations, as in $BaTiO_3$. The magnitude of such spin order induced polarization depends on the displacement of intervening ligand ions ($O^{-2}$) that favour the DM interaction and a net non-zero induced striction. In spin induced ferroelectrics, a non-centrosymmetric magnetic spin order breaks the inversion symmetry, a pre-requisite for generating electric polarization [44]. The increase of magnetic in-equivalence and in-equivalent exchange strictions between the spins order in A and B planes and change of superexchange path lengths ($Fe^{3+}$-$O^{2-}$-$Fe^{3+}$) in rhombohedral structure of Ga doped hematite system is helpful for breaking of inversion symmetry of spin order along off-plane direction. Second source of the breaking of inversion symmetry of spin order could be the magnetically in-equivalent core-shell spin structure in nano-sized grains of Ga doped hematite system (Fig. 11(b)). In a typical AFM system with layered spin structure, the net magnetic moment ($<\mu>$) between two adjacent layers A and B is zero. The net moment is non-zero in the case of canted FM/AFM spin order (Fig. 11(c). In spin canted system, the orientation of magnetic moment vector in the presence of external magnetic field ($H_{ext}$) with respect to local easy axis (EA) is determined by the resultant free energy E = $E_{FM}$+ $E_{AFM}$ + $E_{coupling}$, where $E_{FM}$, $E_{AFM}$, and $E_{coupling}$ represent the free energy terms for FM layer, AFM layer and interlayer spin coupling, respectively [45]. When exchange field corresponds to AFM interactions ($H_{AFM}$) dominates and greater than $H_{ext}$, the in-field magnetization of the system, as we observe at 5 kOe, can relax with time (in-field magnetic relaxation) to achieve a magnetic state near to bulk AFM (core part of the grains); rather than showing an increase of magnetization with time as in the case of a disordered FM [34]. On the other hand, interlayer spin exchange coupling energy ($E_{coupling}$) is largely responsible for electric field induced magnetic exchange bias and magnetic state switching. In the materials like $BiFeO_3$, which is structurally and magnetically similar to Ga doped hematite system, a direct $180^0$ reversal of DM vector by ferroelectric polarization is forbidden from symmetry point of view. However, first-principle calculations [16] predicted a deterministic reversal of the DM vector and canted spin moments by electric field. The coupling between DM vector and induced polarization can switch the magnetization by $180^0$. A complete reversal of magnetization ($180^0$ rotation) by electric field has been predicted in a perpendicularly magnetized thin film [46]. The response of local magnetization vector M(x,t) in spin canted systems, as applicable for the present Ga doped hematite system, to external electric field depends on a spatial distribution of spin magnetization vector components inside the magnetic domains and on the temporal evolution of the magnetization spin vector, determined by Landau-Lifshitz-Gilbert equation: $\frac{\partial \vec{M}}{\partial t} = -\gamma_0 (\vec{M} \times \vec{H}_{eff}) + \frac{\alpha}{M_S} (\vec{M} \times \frac{\partial \vec{M}}{\partial t})$, where $\gamma_0$ is the gyromagnetic ratio and $\alpha$ are the Gilbert damping factor. A competition between the spin torque (first term) and damping torque (second term) determines the orientation of the magnetization vectors either in one side of the interfacial plane or complete reversal about the plane. If we look carefully the magnetization switching in M(H) and M(t) curves of our samples, it is understood that all the spins in the magnetic domains are not participating in spin reversal process under electric field. A fraction (roughly ≤ 1-2 %) of the total spins in a magnetic domain, most probably at the domain walls or shell part of the grains (Fig. 11(e-f)), reverses their magnetic spin directions ($180^0$) upon reversal the polarity of applied electric voltage and rest of the spins (interior to the domain) reluctantly participate in the reversal process. We understand that a unique coupling between magnetic spin order and ferroelectric polarization at the domain walls [36, 41, 43, 47] or core-shell interfaces plays a dominant role in exhibiting magnetic exchange bias effect in Ga doped hematite system. In case of M(H) measurement, core part of the spin structure also contributes in magnetic domain wall motion/domain rotation process. The application of electric field induces a strong magnetic exchange coupling at the interfaces of domain walls or core-shell interfaces. This results in an irreversible shift of the M(H) loop in the presence of electric field. On the other hand, FM coupling among the spins in the presence of magnetic field is stronger than the induced exchange coupling. Hence, the shift of M(H) loop under positive or negative electric voltages is always along positive magnetic field direction with respect to the loop at 0 V. A minor shift of the loop along negative magnetic direction under negative electric field in comparison to the loop under positive electric field confirms that only a fraction of interfacial spins participates in electric field controlled 180 $^{0}$ reversal process. A complete 180 $^{0}$ reversal (switching) of these interfacial magnetic spin vectors by reversing the electric field polarity is macroscopically observed in Ga doped hematite system during M(t) measurement with reference to the magnetization state at 0 V, although the net magnetization is always positive under positive 5 kOe magnetic field and magnetic state switches to higher level and lower level for application of positive and negative electric voltage, respectively.

## 5. Conclusions

The replacement of magnetic Fe atoms by non-magnetic Ga atoms in rhombohedral structure of $\alpha$-$Fe_2O_3$ enhances magnetoelectric properties and magnetic order has been diluted at considerably higher value of Ga content. The ferromagnetic loop, being a universal feature irrespective of metal, semiconductor or insulator, under external electric field was used to detect the existence of magneto-electric coupling in Ga doped hematite system. The Ga doped hematite system exhibited similar lattice (rhombohedral) and magnetic (spin canting) structure as in $BiFeO_3$ and both are lead free system. The magneto-electric coupling in Ga doped hematite system (rhombohedral structure) is free from strain mediated coupling effect and considered to be intrinsic of the material. The magnetic exchange bias shift, generally a low temperature phenomenon in magnetically bi-layered (FM/AFM) systems after high magnetic field cooling from higher temperature, has been observed in Ga doped hematite system at room temperature itself without magnetic field cooling and controlled absolutely by electric field. A complete reversal of the magnetization vectors (negative to positive or vice versa) is not observed at macroscopic level upon reversing the polarity of electric voltage during in-field magnetic relaxation of spins, but the samples instantly jumped to higher magnetic state at the time of applying positive voltage ($\Delta M \sim$ 2-2.5 %) and to lower magnetic state at the time of applying negative voltage. The electric field controlled magnetic exchange bias and switching behaviour can be understood in terms of spin induced ferroelectricity /multiferroic properties at the domain walls or exchange coupling at the interfaces of core-shell spin structure. It is understood that a fraction (roughly $\leq$ 1-2 %) of the total spins at the domain walls or core-shell interface reverses their magnetic spin directions (180 $^0$) upon reversal the polarity of applied electric voltage and rest of the spins (interior to the domain or core) reluctantly participate in the reversal process. A complete 180 $^0$ reversal (switching) of the interfacial magnetic spin vectors by reversing the electric field polarity is macroscopically confirmed during M(t) measurement with reference to the magnetization state at 0 V. The electric field controlled magnetic state in Ga doped hematite system at room temperature can have a tremendous impact on developing spintronics based technology, non-volatile memory and electro-magnetic switching devices.

**Acknowledgment**

Authors thank to CIF, Pondicherry University for providing some experimental facilities. RNB acknowledges research grants from Department of Science and Technology (NO.

**Figure captions**

Fig. 1 Profile fitting of XRD pattern (left hand side) and Micro-Raman spectra (righ hand side) confirm rhombohedral structure in Ga doped hematite system.

Fig. 2 (a) M(H) loops of hematite sample measured at different electric voltages, magnetic relaxation at 5 kOe for 0 V (b), for ON-OFF modes of +100 V (c), for ON-OFF modes of -100 V.

Fig.3 M(H) loops at different measurement voltages (a), magnified M(H) loops at 0V, 200 V (b), variation of magnetic exchange bias field with applied voltages (c). Time dependence of magnetic moment at a magnetic field of 5 kOe and different voltage conditions 0V and +100V (d), 0V and -100V (e), and 0V - $\pm$ 500V in steps of 100V (f).

Fig. 4(a) Room temperature M(H) loops at different measurement voltages, (b) magnified M(H) loops, (c) M(H) loops at 0 V and $\pm$ 200 V for Ga04MA100V8 sample, (d) variation of exchange bias with different positive applied voltages.

Fig. 5 Magnified form of room temperature M(H) loops of Ga04MA50V8 sample, measured at different electric voltages (a), shown for +200 V and -200 V with respect to 0 V loop (b). Insets show the variation of magnetic exchange bias field for different electric voltage (a) and for electric bias voltages at $\pm$100 V and $\pm$ 200 V (b).

Fig. 6 Room temperature time dependent magnetic moment at a magnetic field of 5 kOe and different voltage conditions (a) 0V, (b) 0V and 100V, (c) 0V and +100V, and (d) 0V and -100V, (e) 0V and +200V, (f) 0V and -200V for Ga04MA100V8 sample.

Fig. 7 Time dependent magnetic moment at a magnetic field of 5 kOe and voltage conditions (a) 0V and +100V, (b) 0V and -100V, (c) 0V and 100 V, (d) branch of (a), (e) branch of (b), (f) magnetization at voltage 0-500 V, (g) change of magnetization.

Fig. 8  Room temperature M(H) loops at different positive voltages (a) and a comparative M(H) loops at $\pm$ 200 V and at 0 V (b). The Inset shows variation of exchange bias field and coercivity at different voltages for Ga08MA25V8 sample.

Fig.9 In field magnetic relaxation at magnetic field of 5 kOe for applied electric voltage at 0 V (a). The in-field magnetic relaxation at ON-OFF mode of voltage for +200 V (b), for -200V (c), for $\pm$200V (d), and increment of voltage up to $\pm$ 500V in steps of 100V (e). The charge of switched magnetization at peak values with applied electric field (f) is shown for Ga08MA25V8 sample.

Fig. 10 M(H) loops for Ga10MM50V7 sample measured at different electric voltages (a) and loop shift is shown in magnified plot (b). Variation of the magnetic coercivity and exchange bias shift at 200 V with Ga content in the Ga doped hematite system (c).

Fig. 11 A schematic diagram of the spin order between two planes in $\alpha$-$Fe_2O_3$ and Ga doped $\alpha$-$Fe_2O_3$ (a), Core-shell spin structure in a grain of Ga doped $\alpha$-$Fe_2O_3$ (b), in-field magnetic relaxation at V = 0 (c) and in the presence of constant voltage (d), response of spin vectors during M(H) measurement in the presence of constant +ve (e) and −ve (f) voltage.

Table 1. Structural information of the samples used in the present study

| Sample code | Structural formula | Preparation condition | Cell parameters a(±0.005Å), c(±0.002Å), V (±0.05(Å)$^3$ | | | Grain size (nm) |
|---|---|---|---|---|---|---|
| | | | $a$ (Å) | $c$ (Å) | $V$(Å$^3$) | |
| Ga02MA100V8 | $\alpha$-$Fe_{1.8}Ga_{0.2}O_3$ | Mechanical alloyed for 100 h and annealed at 800 $^\circ$C under vacuum for 6 h | 5.033 | 13.738 | 301.41 | 53 |
| Ga04MA50V8 | $\alpha$-$Fe_{1.6}Ga_{0.4}O_3$ | Mechanical alloyed for 50 h and annealed at 800 $^\circ$C under vacuum for 2 h | 5.036 | 13.744 | 301.99 | 37 |
| Ga04MA100V8 | $\alpha$-$Fe_{1.6}Ga_{0.4}O_3$ | Mechanical alloyed for 100 h and annealed at 800 $^\circ$C under vacuum for 2 h | 5.035 | 13.715 | 301.15 | 26 |
| Ga08MA25V8 | $\alpha$-$Fe_{1.2}Ga_{0.8}O_3$ | Mechanical alloyed for 25 h and annealed at 800 $^\circ$C under vacuum for 2 h | 5.025 | 13.703 | 299.70 | 50 |
| Ga10MM50V7 | $\alpha$-$FeGaO_3$ | Mechanical alloyed for 100 h, followed by heating at 1250 $^\circ$C in air. This is following mechanical milling for 50 h and subsequent vacuum annealing at 700 $^0$C (2 h) | 5.024 | 13.611 | 297.56 | 19 |

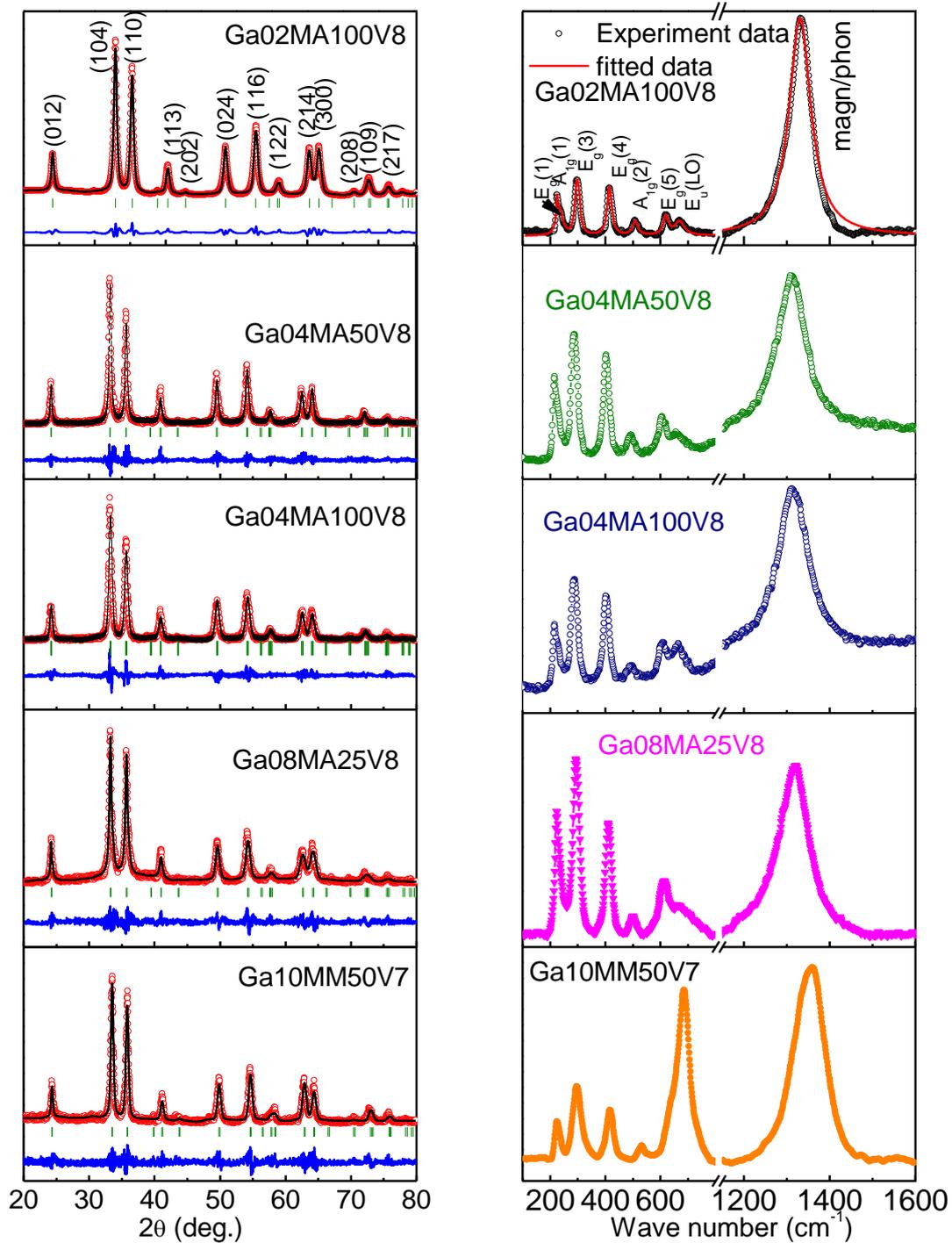

Fig. 1 Profile fitting of XRD pattern (left hand side) and Micro-Raman spectra (righ hand side) confirm rhombohedral structure in Ga doped hematite system.

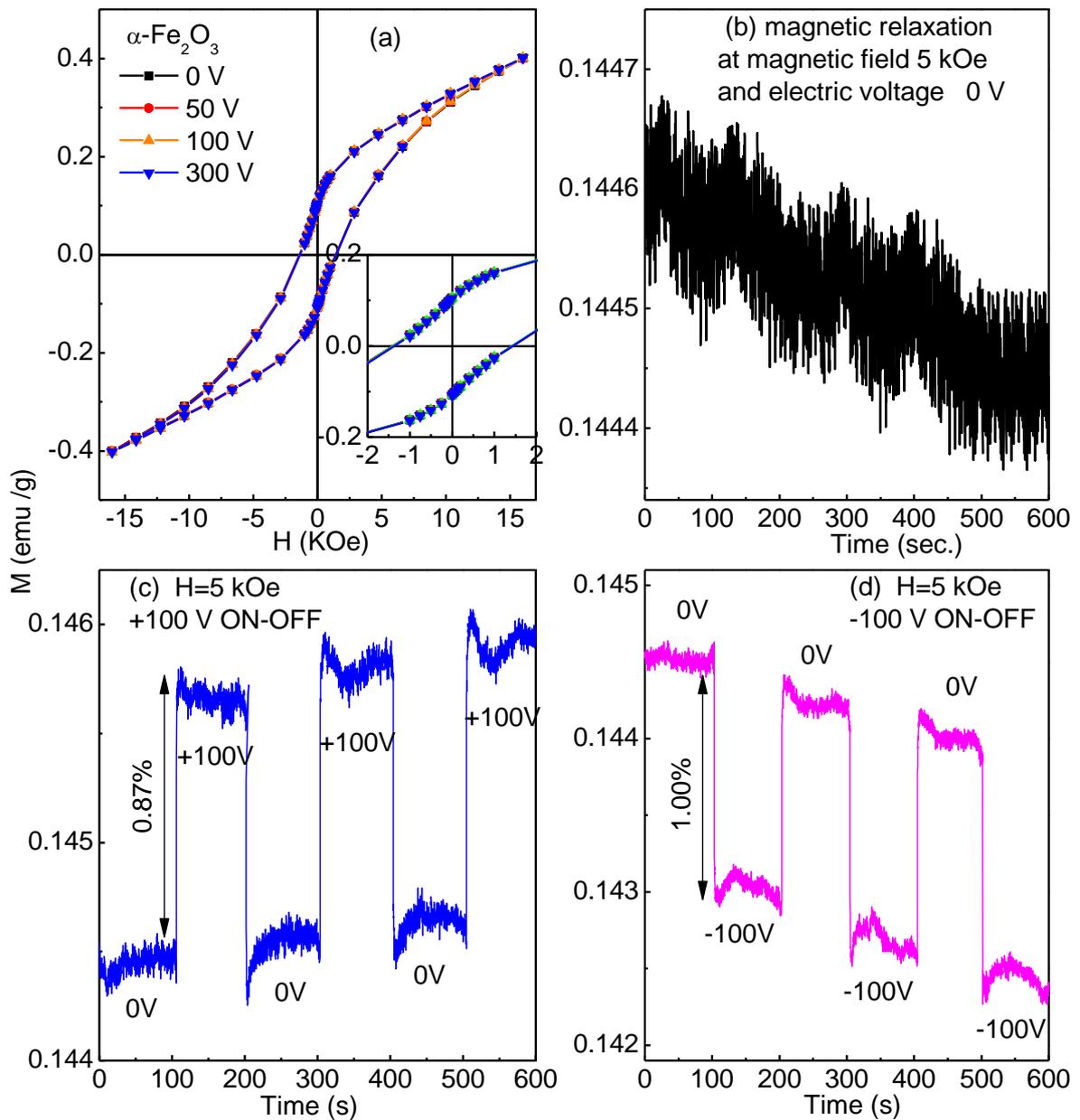

Fig. 2 (a) M(H) loops of hematite sample measured at different electric voltages, magnetic relaxation at 5 kOe for 0 V (b), for ON-OFF modes of +100 V (c), for ON-OFF modes of -100 V.

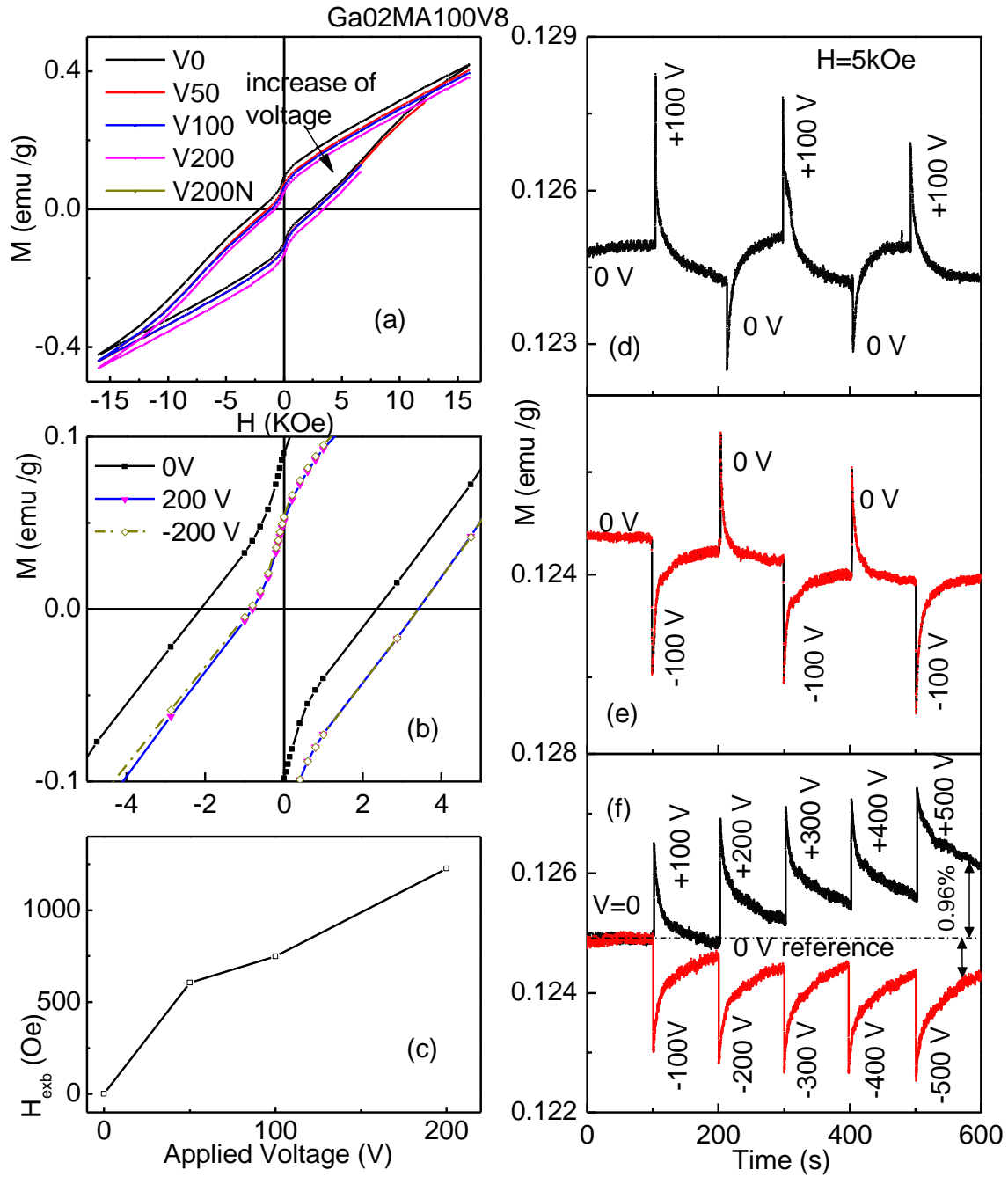

Fig.3 M(H) loops at different measurement voltages (a), magnified M(H) loops at 0V, 200V (b), variation of magnetic exchange bias field with applied voltages (c). Time dependence of magnetic moment at a magnetic field of 5 kOe and different voltage conditions 0V and +100V (d), 0V and −100V (e), and 0V - ± 500V in steps of 100V (f).

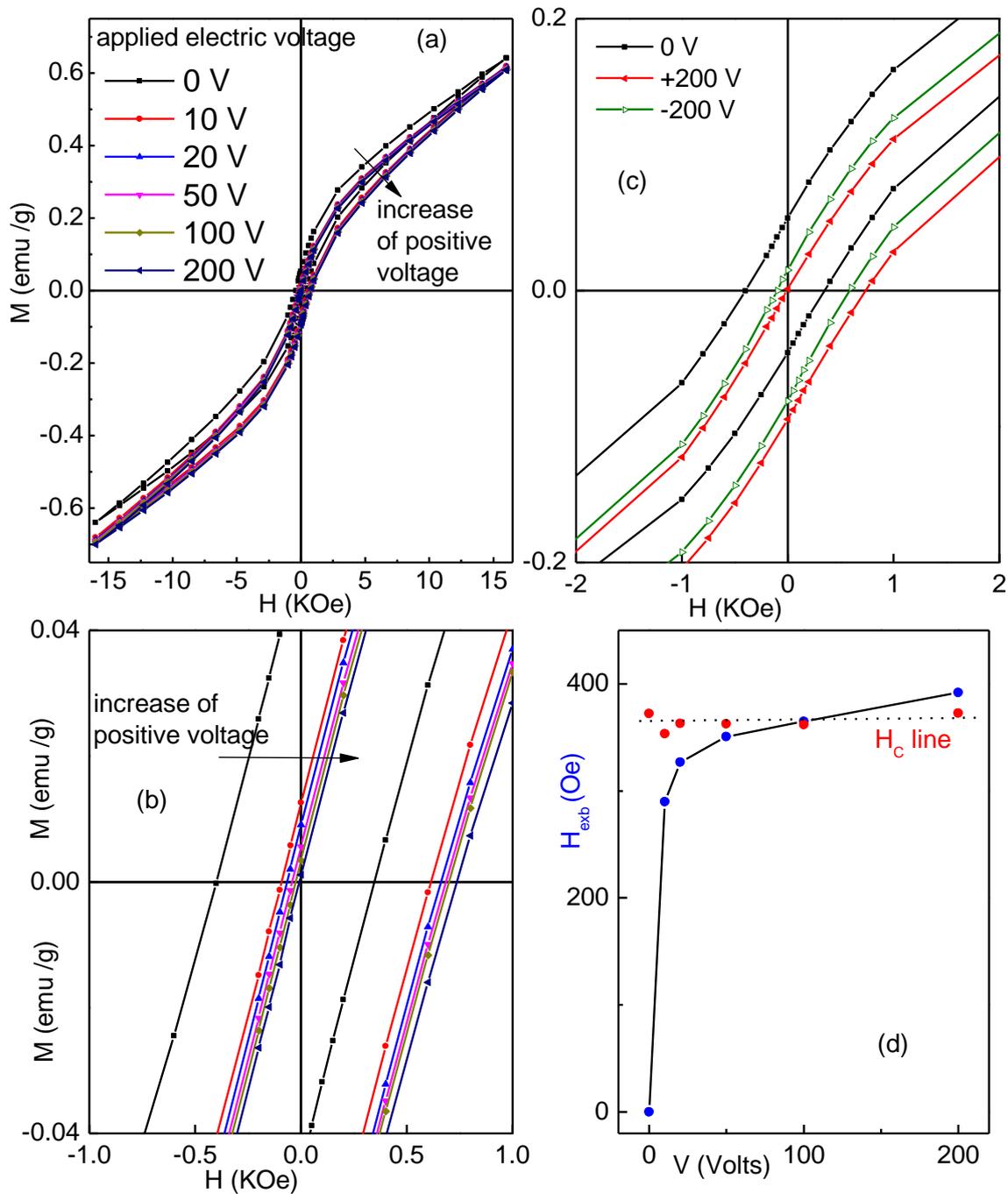

Fig. 4(a) Room temperature M(H) loops at different measurement voltages, (b) magnified M(H) loops, (c) M(H) loops at 0 V and ± 200 V for Ga04MA100V8 sample, (d) variation of exchange bias with different positive applied voltages.

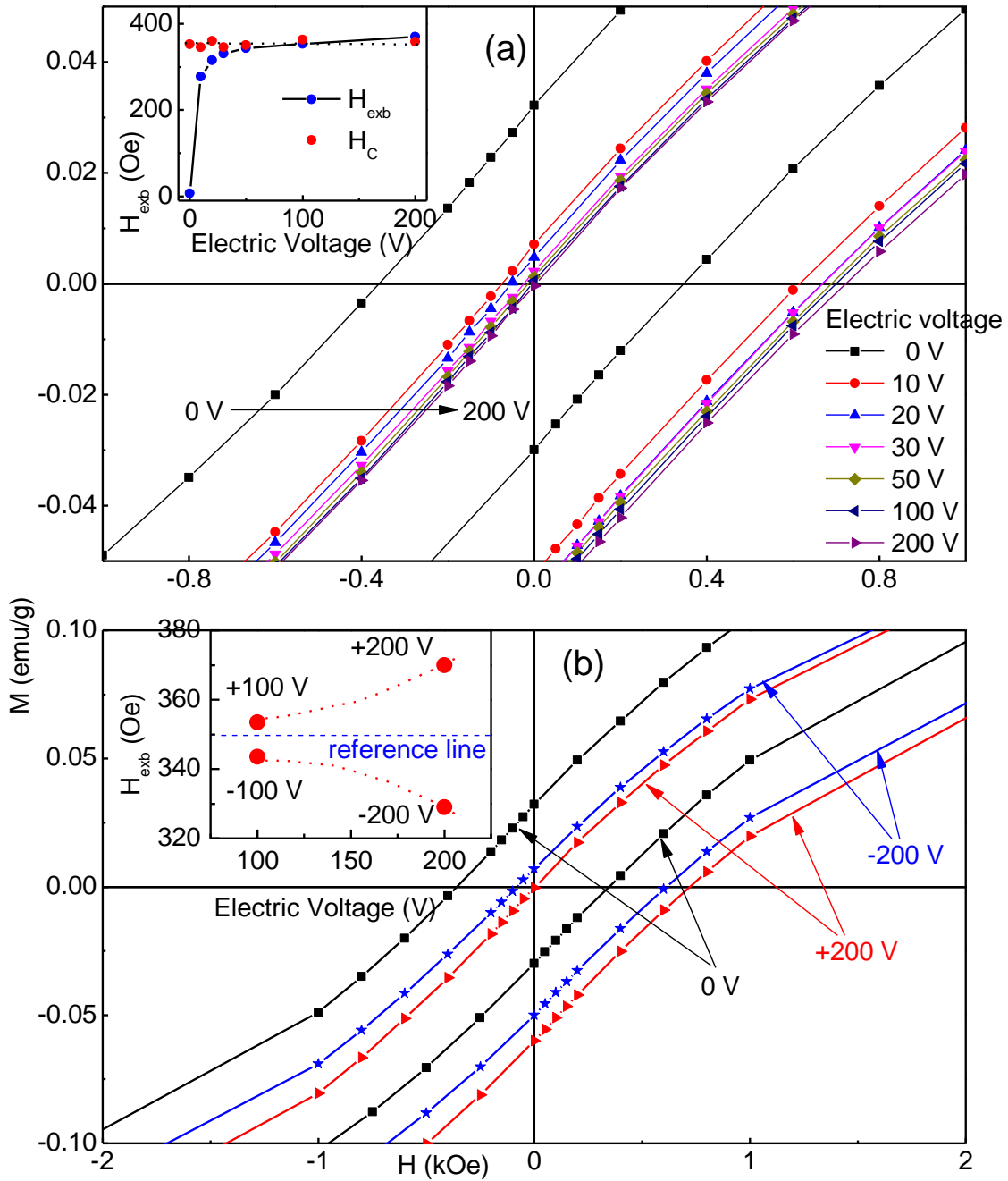

Fig. 5 Magnified form of room temperature M(H) loops of Ga04MA50V8 sample, measured at different electric voltages (a), shown for +200 V and -200 V with respect to 0 V loop (b). Insets show the variation of magnetic exchange bias field for different electric voltage (a) and for electric bias voltages at $\pm$100 V and $\pm$200 V (b).

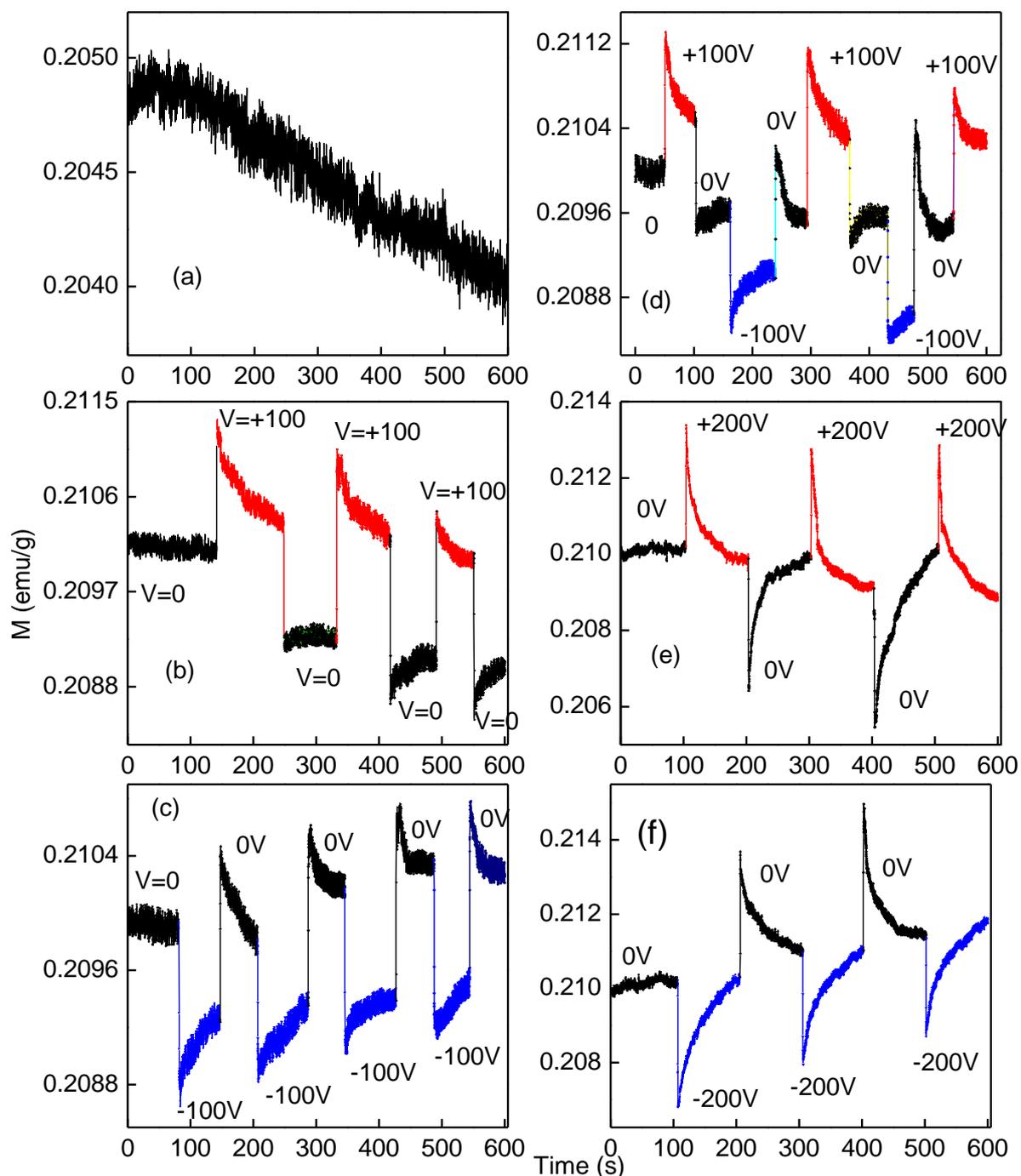

Fig. 6 Room temperature time dependent magnetic moment at a magnetic field of 5 kOe and different voltage conditions (a) 0V, (b) 0V and ±100V, (c) 0V and +100V, and (d) 0V and -100V, (e) 0V and +200V, (f) 0V and -200V for Ga04MA100V8 sample.

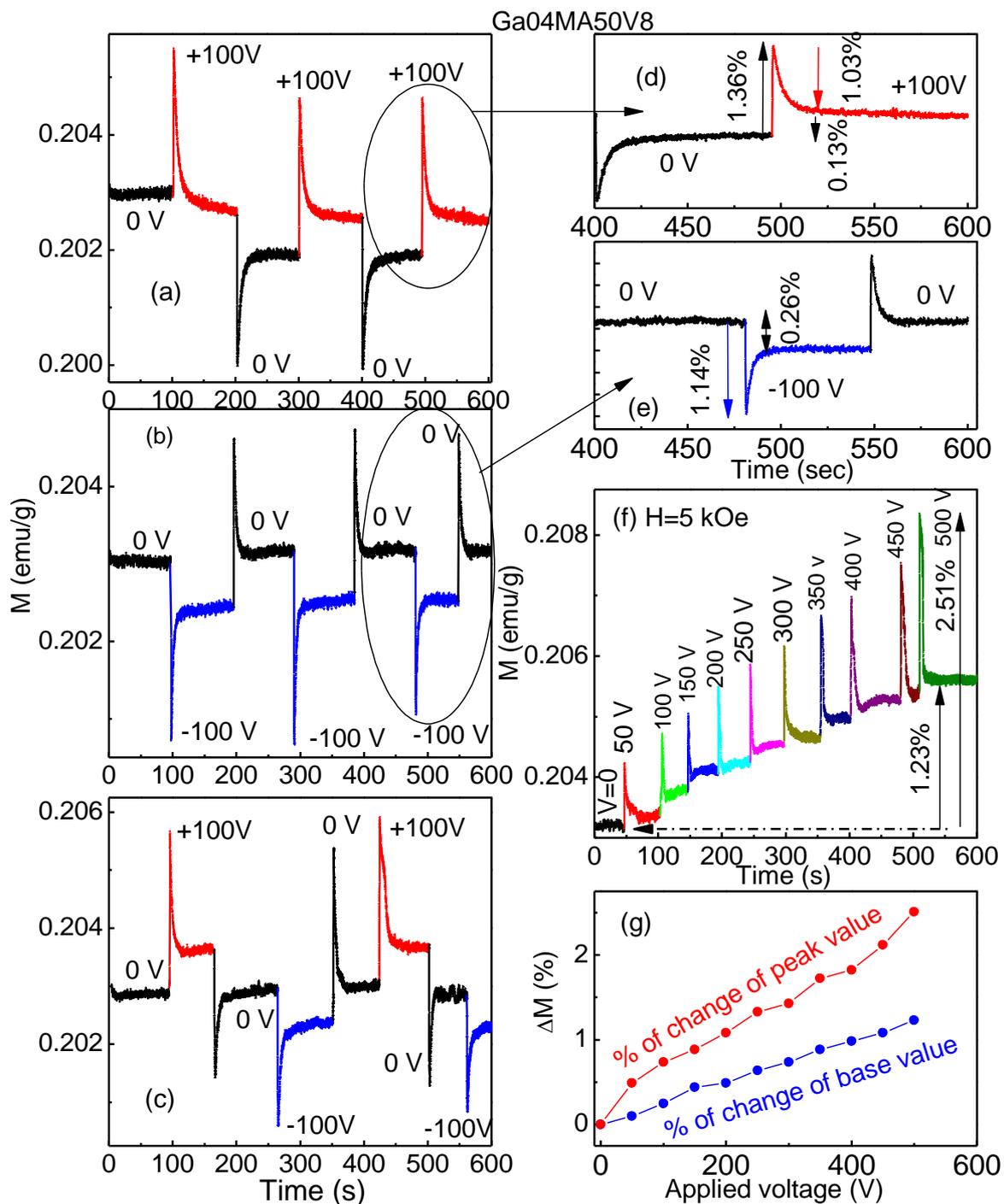

Ga04MA50V8

Fig. 7 Time dependent magnetic moment at a magnetic field of 5 kOe and voltage conditions (a) 0V and +100V, (b) 0V and −100V, (c) 0V and ±100 V, (d) branch of (a), (e) branch of (b), (f) magnetization at voltage 0-500 V, (g) change of magnetization.

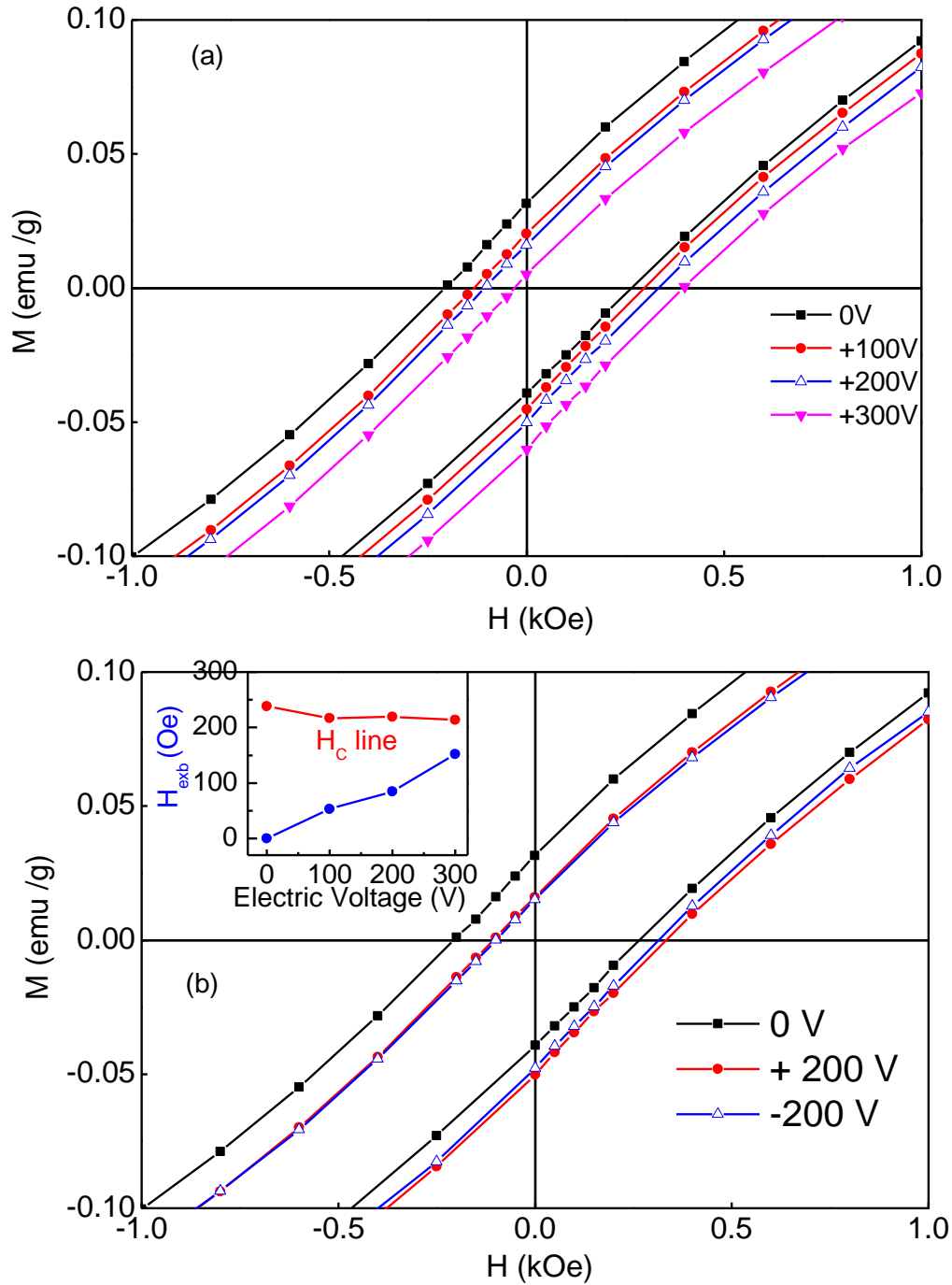

Fig. 8  Room temperature M(H) loops at different positive voltages (a) and a comparative M(H) loops at ± 200 V and at 0 V (b). The Inset shows variation of exchange bias field and coercivity at different voltages for Ga08MA25V8 sample.

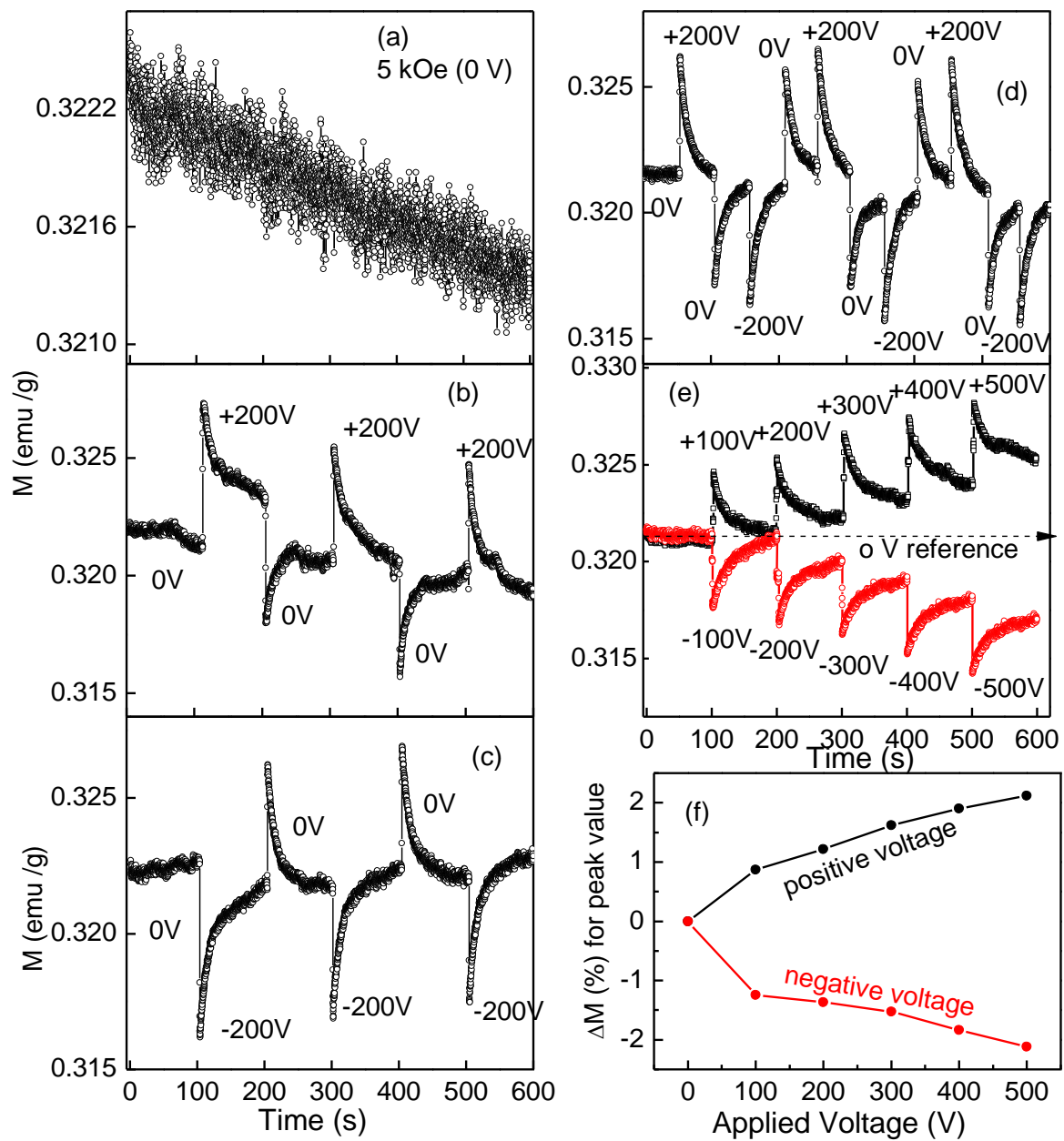

Fig.9 In field magnetic relaxation at magnetic field of 5 kOe for applied electric voltage at 0 V (a). The in-field magnetic relaxation at ON-OFF mode of voltage for +200 V (b), for −200V (c), for ± 200V (d), and increment of voltage up to ± 500V in steps of 100V (e). The charge of switched magnetization at peak values with applied electric field (f) is shown for Ga08MA25V8 sample.

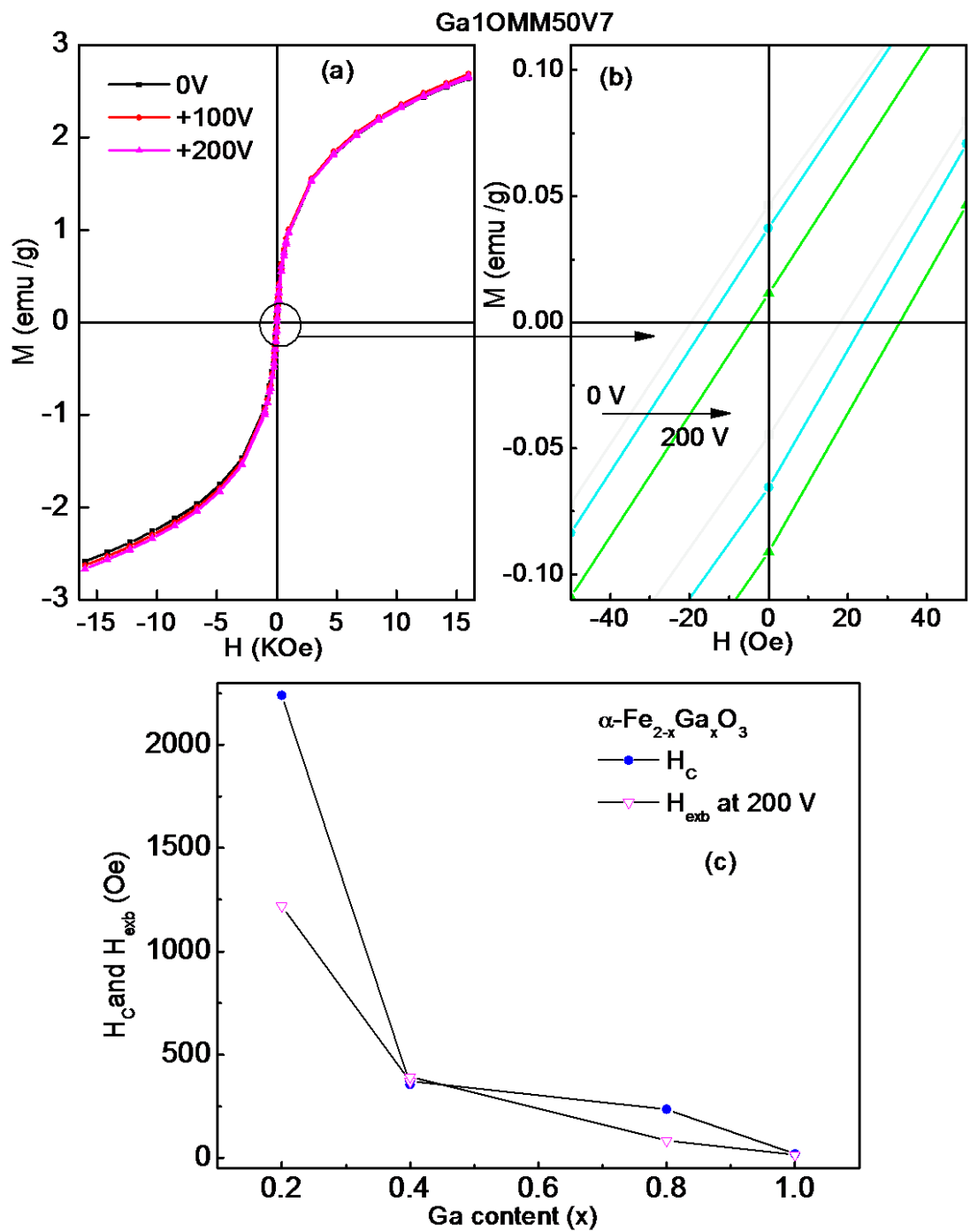

Fig. 10 M(H) loops for Ga10MM50V7 sample measured at different electric voltages (a) and loop shift is shown in magnified plot (b). Variation of the magnetic coercivity and exchange bias shift at 200 V with Ga content in the Ga doped hematite system (c).

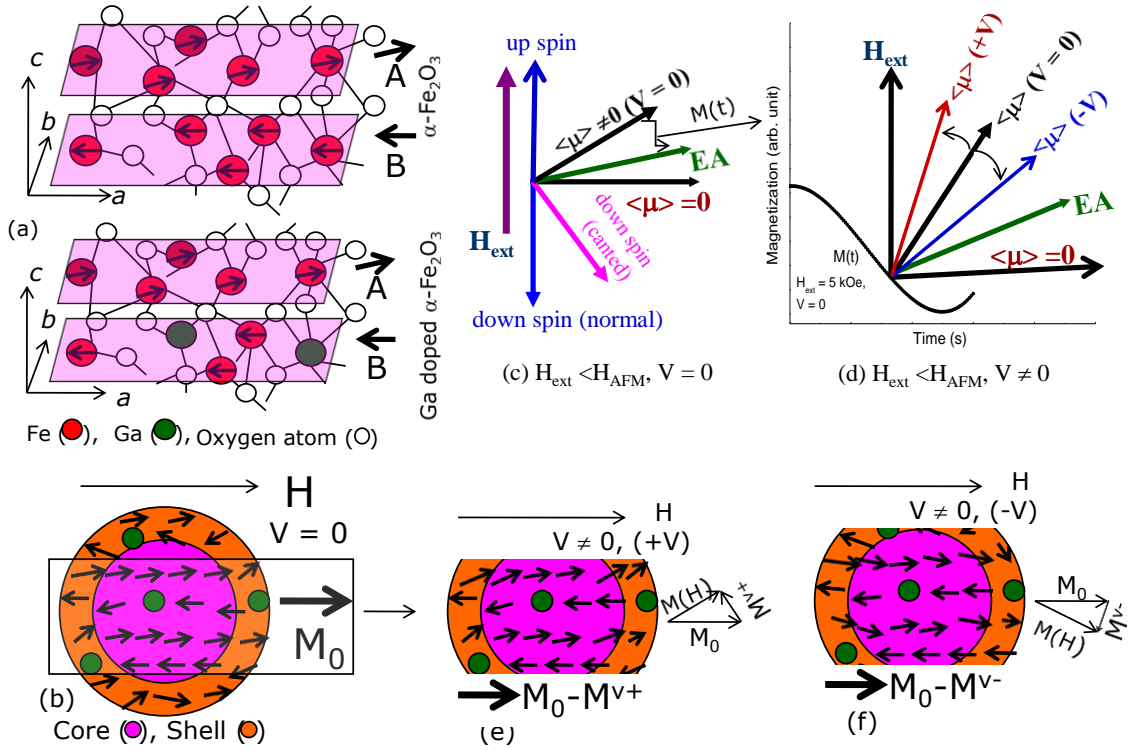

Fe ●, Ga ●, Oxygen atom (O) ○

Core ●, Shell ●

Fig. 11A schematic diagram of the spin order between two planes in α-Fe$_2$O$_3$ and Ga doped α-Fe$_2$O$_3$ (a), Core-shell spin structure in a grain of Ga doped α-Fe$_2$O$_3$ (b), in-field magnetic relaxation at V = 0 (c) and in the presence of constant voltage (d), response of spin vectors during M(H) measurement in the presence of constant +ve (e) and −ve (f) voltage.